\documentclass[letterpaper]{article} 
\usepackage[submission]{aaai23}  
\usepackage{times}  
\usepackage{helvet}  
\usepackage{courier}  
\usepackage[hyphens]{url}  
\usepackage{graphicx} 
\usepackage{natbib}  
\usepackage{caption} 
\usepackage{algorithm}
\usepackage{algorithmic}

\usepackage{newfloat}
\usepackage{listings}
\DeclareCaptionStyle{ruled}{labelfont=normalfont,labelsep=colon,strut=off} 
\floatstyle{ruled}
\newfloat{listing}{tb}{lst}{}
\floatname{listing}{Listing}
\usepackage{siunitx}

\title{Learning Affects Trust: Design Recommendations and Concepts for Teaching Children---and Nearly Anyone---about Conversational Agents}
\author {
    Jessica Van Brummelen,\textsuperscript{\rm 1}
    Mingyan Claire Tian,\textsuperscript{\rm 1}
    Maura Kelleher,\textsuperscript{\rm 2}
    Nghi Hoang Nguyen,\textsuperscript{\rm 1}
}
\affiliations {
    \textsuperscript{\rm 1} Massachusetts Institute of Technology\\
    jess@csail.mit.edu, maurakel@mit.edu, nghin@mit.edu
    
    \textsuperscript{\rm 2} Wellesley College\\
    mt1@wellesley.edu
}

\usepackage{bibentry}

\begin{document}

\maketitle

\begin{figure*}[ht!]
       \centering
	\includegraphics[width=.95\linewidth]{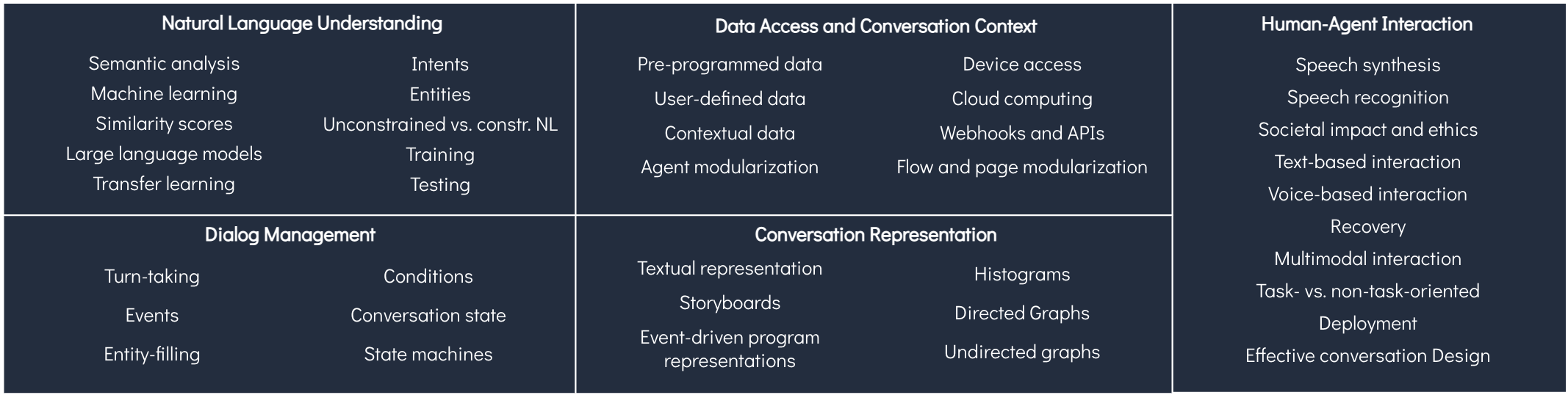}
	\caption{The framework of forty CA concepts, which are described in-depth in the Appendix \protect{\cite{concepts-to-teach-conv-ai-appendix}}.}
	\label{fig:agent-concepts}
\end{figure*}

\begin{abstract}

Research has shown that human-agent relationships form in similar ways to human-human relationships. Since children do not have the same critical analysis skills as adults (and may over-trust technology, for example), this relationship-formation is concerning. Nonetheless, little research investigates children’s perceptions of conversational agents in-depth, and even less investigates how education might change these perceptions. We present K-12 workshops with associated conversational AI concepts to encourage healthier understanding and relationships with agents. Through studies with the curriculum, and children and parents from various countries, we found participants' perceptions of agents---specifically their partner models and trust---changed. When participants discussed changes in trust of agents, we found they most often mentioned learning something. For example, they frequently mentioned learning where agents obtained information, what agents do with this information and how agents are programmed. Based on the results, we developed recommendations for teaching conversational agent concepts, including emphasizing the concepts students found most challenging, like training, turn-taking and terminology; supplementing agent development activities with related learning activities; fostering appropriate levels of trust towards agents; and fostering accurate partner models of agents. Through such pedagogy, students can learn to better understand conversational AI and what it means to have it in the world.
\end{abstract}

\section{Introduction and Related Work}

Reports have indicated an exponential rise in the use of voice-based agents, like Alexa and Siri \cite{exponential-rise-voice-agents-report,voice-based-on-frontline}. Researchers have taken note and begun to investigate the societal implications of agent ubiquity, like how this may affect social norms, decision-making and information spread \cite{trust-anthropomorphism-relationships-cas,algorithmic-aversion-clinical-deployment}. For instance, researchers have found relationships with agents form similarly to human-human relationships %
\cite{trust-anthropomorphism-relationships-cas,transparency-relationship-development}. One concern researchers have %
is how humans may over-trust agents, especially if they have personified traits \cite{agent-personality-trust,alexa-perceptions-idc}. Considering how trust is connected to relationship building, and is a key factor in misinformation spread \cite{misinformation-trust-conspiracy-beliefs,trust-anthropomorphism-relationships-cas}, it is important for people to be able to calibrate their levels of trust towards conversational agents (CAs), according to agents' actual trustworthiness. In this paper, we investigate the link between a conversational AI educational intervention and participants' perceptions and levels of trust of CAs, finding that participants most often mentioned learning something about CAs as reasons for changes in their trust.

Along similar lines, many researchers have started AI education initiatives. For instance, the AI4K12 initiative is developing tools and curriculum based on five core ``Big AI Ideas''; %
MIT's %
RAISE initiative is developing vocational and K-12 tools for AI education; and Code.org is developing interactive resources for K-12 AI education \cite{big5,raise-mit,code-org}. Nonetheless, very few resources specifically teach about conversational AI or investigate how to best do so, despite researchers noting the importance of teaching a breadth of types of AI, a need for CA pedagogical artifacts, and the unique societal questions and challenges CAs present due to their relational nature \cite{ailiteracy,vui-curriculum-review,trust-anthropomorphism-relationships-cas}. There are especially few resources in the K-12 space, despite children potentially being more vulnerable to %
misinformation spread \cite{vui-curriculum-review,teaching-tech-talk-eaai}.

One notable K-12 resource for CA education includes \citet{daniella-thesis-child-robot-relationships}'s social robotics curriculum. In this curriculum, students aged 9-12 learn about the societal impact of social robots and how to prototype robot conversation. %
It focuses on social robotics and includes a portion on conversational AI, in which students learn concepts including conversational flow representation and machine learning (ML). Another resource includes \citet{convo-ml}'s CA curriculum, in which students aged 13-15 develop CAs through speaking with \textsc{Convo}, and learn concepts like training ML models and the difference between constrained and unconstrained natural language (NL). A third resource includes \citet{teaching-tech-talk-eaai}'s conversational AI curriculum. It teaches students aged 13-18 how to create CAs using a block-based coding interface, and related concepts like intent-modeling and entity-filling. %
Since this curriculum specifically focuses on conversational AI, has a broad intended age range and a low-barrier-to-entry, open-source interface for developing CAs, we build on it to investigate our research questions. %
We refer to the interface as ``ConvoBlocks''.

\subsection{Study Novelty}
With this interface, \citet{alexa-perceptions-idc} investigated changes in students' perceptions of CAs through agent-building workshops. They found correlations between perceptions of Alexa's friendliness, safeness, and trustworthiness. Trustworthiness as a concept, however, is very broad, and no one (to our knowledge) has investigated how learning to program CAs affects specific aspects of trust. %
Understanding this, however, could help educators better develop pedagogy to empower students to understand agents' true trustworthiness. %
In our study, we adopt the widely-used model of trust by \citet{trust-e-commerce-typology-model}. %
This model consists of four characteristics of trust, (1) \textbf{competence}, (2) \textbf{benevolence}, (3) \textbf{integrity} and (4) \textbf{predictability}. %
We also investigate people's trust of agents' correctness, as this relates directly to misinformation spread.

Another construct that could help improve agent pedagogy includes ``partner models'', which define how users perceive their conversational partners, or in this case, CAs. \citet{partner-models-doyle}'s model %
involves three dimensions: (1) \textbf{competence and dependability}, (2) \textbf{human-likeness}, and (3) \textbf{cognitive flexibility}. By understanding users' partner models of CAs, we can better understand their expectations and reactions. %
For instance, if a user expects an agent to be flexible, but it is not (e.g., it only understands very specific commands), the user may become frustrated. However, by understanding users' partner models, CA designers can develop agents to foster accurate partner models (e.g., an agent which outlines the extent of its flexibility) \cite{partner-modelling-cowan}. In a similar way, %
by understanding students' partner models, educators can develop pedagogy to empower students to develop more accurate perceptions of agents, allowing them to better understand how they work.

There is also little literature investigating the perceptions of people from countries that are not Western, Educated, Industrialized, Rich and Democratic (WEIRD) \cite{weird-hci-chi}, and to the authors' knowledge, no literature investigating the difference between how child and parent perceptions of CAs change after programming them. Thus, we incorporate these groups of people in our study %
and aim towards developing teaching recommendations for nearly anyone to learn about CAs (although we realize there is still much work to be done in this area, and look forward to other researchers continuing in this vein). %

\subsection{Other Educational Interventions Affecting Trust}
With other technologies, researchers have shown how educational interventions can affect trust, increase understanding and decrease the spread of misinformation \cite{misinformation-education-conspiracy,misinformation-education-twitter,misinformation-trust-ml-warnings,transparency-relationship-development}. In terms of investigating changes in children's trust of technology, \citet{daniella-thesis-child-robot-relationships} found children trusted robots less after engaging in societal impact curriculum. Others found children's trust decreased after learning about the programmatic nature of robots \cite{transparency-relationship-development}. To the authors' knowledge, the only educational intervention in which researchers have investigated changes in children's trust of CAs is the ConvoBlocks study discussed above. \citet{alexa-perceptions-idc} did not find any significant differences in students' perceptions of agents' trustworthiness through the workshops; however, they did find correlations between perceptions of trustworthiness, safeness and friendliness. Our study investigates whether there are changes in trust for specific subsets of participants not studied previously (e.g., children vs. parents, WEIRD vs. non-WEIRD).

\subsection{Conversational Agent Concepts}
As mentioned previously, there is a lack of CA-specific pedagogical materials, despite conversational AI's unique positioning in terms of market penetration and potential to be a primary mode of human-computer interaction \cite{vui-curriculum-review,voice-commerce-stats,voice-hci-review}. To begin, there exists a need for a framework of CA-specific competencies, such that educators can develop comprehensive curricula, teaching the many aspects of CA development and design. To address this need, we developed a framework of forty CA concepts, as presented in the thesis, \citet{vanbrummelen-phd}, and in Figure \ref{fig:agent-concepts}. We developed our workshop curriculum based on teaching a number of these concepts, as described in the Procedure section. To develop this framework, we analyzed %
CA development platforms, CA educational materials and CAs %
with intended audiences from primary-school-aged students to advanced software developers. %

\subsection{Research Questions}
To address the above literature gaps, we investigated the following research questions through our CA programming and societal impact educational intervention: %

%
%
%

%
    \textbf{RQ1:} What do various people (WEIRD, non-WEIRD, and different generations) find %
    difficult in the intervention? %

    \textbf{RQ2:} How do various people feel in terms of self-efficacy and programming identity through the intervention? %

    \textbf{RQ3:} How do various people perceive Alexa with respect to partner models and trust through the intervention? %

    \textbf{RQ4:} How might the results from RQ1-3 inform teaching guidelines for conversational AI?

Through investigating these research questions, we developed design recommendations (DRs) for  CA pedagogy. Analysis of our results also led to agent DRs, which are examined in another paper, \citet{want-perceive-trust-transparency-chi}.

\section{Participants}
There were 99 pairs of children and parents who filled the interest forms to participate in the workshops. In total, 49 completed at least 1 of the 3 surveys. %
There were 27 children (age avg.=13.96, SD=1.829) and 19 parents (age avg.=46.35, SD=11.07) on the pre-survey. From the same survey, 23 participants were from WEIRD countries (age avg.=26.45, SD=19.24) and 23 were from non-WEIRD countries (age avg.=25.48, SD=15.18). We defined WEIRD countries according to \citet{weirdness-calc}'s method. Participants were from the US, Singapore, Canada and New Zealand (WEIRD, 57\% from the US); and Indonesia, Iran, Japan, and India (non-WEIRD, 87\% from Indonesia).

\section{The ConvoBlocks Interface}
\begin{figure}[htb!]
    \centering
	\includegraphics[width=\linewidth]{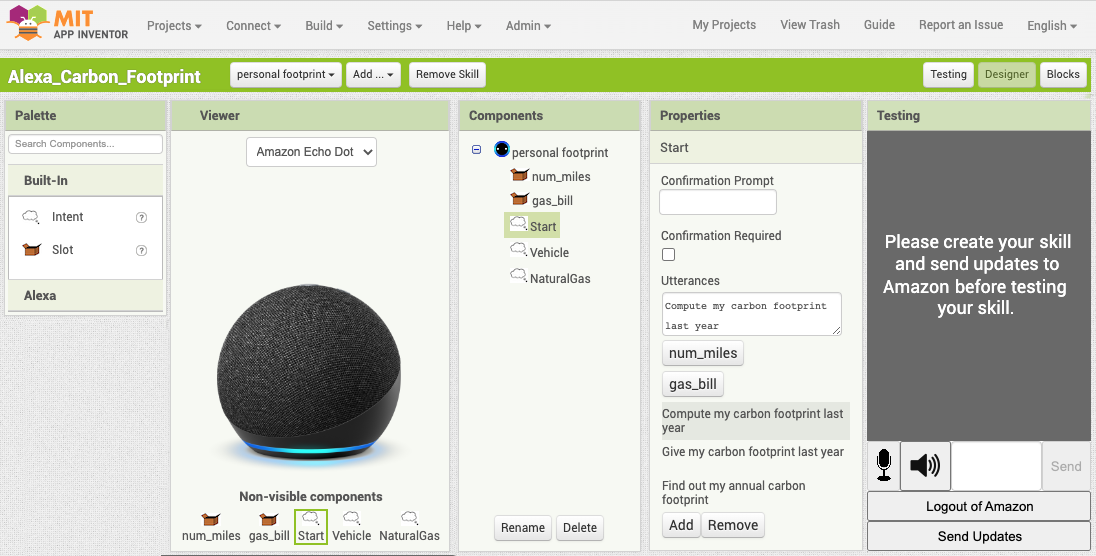}
	\caption{The ConvoBlocks page in which users can define intents and entities for their agents to recognize.}
	\label{fig:convoblocks-interface}
\end{figure}

\begin{figure}[htb!]
    \centering
	\includegraphics[width=\linewidth]{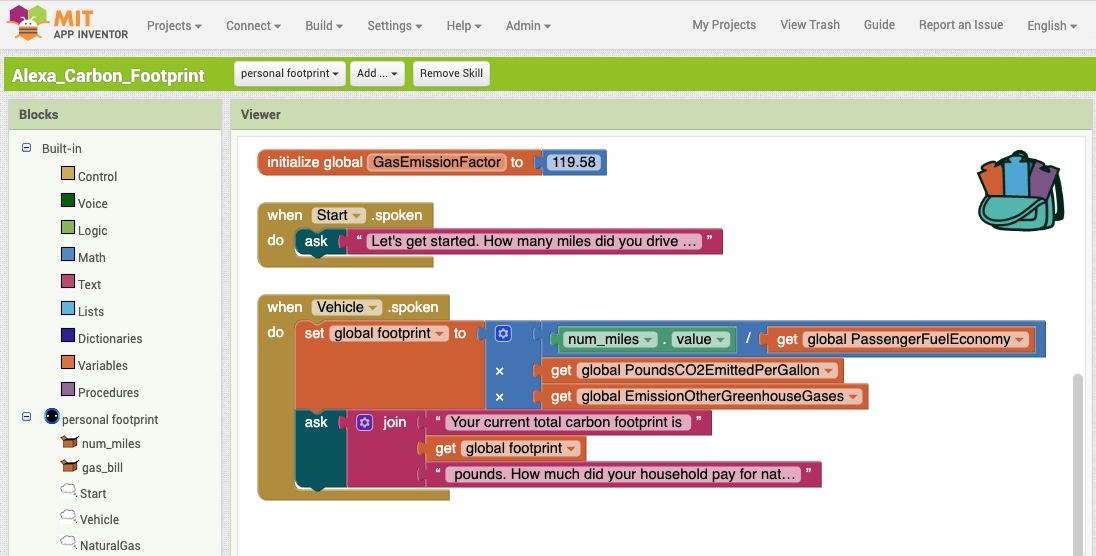}
	\caption{The ConvoBlocks page in which users can program agents to respond to intents.}
	\label{fig:convoblocks-blocks}
\end{figure}

We utilized ConvoBlocks (aka the MIT App Inventor Conversational AI Interface) to teach participants how to program CAs due to its low barrier to entry, its wide target age range, how it is open-source and how participants can create CAs to run on real smart-home devices, like Amazon Echoes \cite{teaching-tech-talk-eaai}. To create such agents, participants go to a web page where they can define CA invocation names (e.g., ``Carbon Footprint Calculator''), intents (e.g., ``What's my carbon footprint?'') and entities (e.g., miles driven), as in Figure \ref{fig:convoblocks-interface}. Next, they go to a page where they can connect code blocks to define how the CA responds to particular intents (e.g., by saying ``Your footprint is 11.3 tons per year''), as in Figure \ref{fig:convoblocks-blocks}. Participants can test their agent on the web page itself or any Alexa-enabled devices, like an Alexa App or Amazon Echo.

\section{Procedure}
The workshops spanned two days virtually over Zoom for approximately 3.5 hours each day. %
The first day introduced CA concepts through two agent coding tutorials. Participants created agents to calculate carbon footprints. We gave participants PDFs of these two tutorials, plus a third, to be completed if they finished early. %
After the tutorials, participants listed traits of their ``ideal'' CAs in an ideation session.

Prior to the tutorials, participants completed a pre-survey with questions about demographics and their perceptions and trust of CAs. At the end of Day 1, they filled out a mid-survey similar to the pre-survey. It also asked whether and why their responses changed, and which CA concepts were most challenging. On Day 1 we focused on teaching participants the CA concepts, \textit{Training}, \textit{Intents}, \textit{Agent modularization}, \textit{Entities}, \textit{Events}, \textit{Testing}, \textit{Turn-taking} and associated CA terminology. These were mentioned on the mid-survey. %

The second day focused on teaching %
\textit{Societal impact and ethics} through presentations and group activities. %
Instructors of the workshop presented on current world challenges. Participants %
discussed sustainability and how technology---including conversational AI---may influence human mindsets. They ultimately created presentations on how technology could be used to combat sustainability problems. A final survey was given after the presentations, which again asked participants to reflect on how their partner models and trust of agents changed during the workshops.

\section{Data analysis}

We analyzed the long-answer responses and ideation data %
using a coding reliability approach to thematic analysis, as described by \citet{thematic-analysis-open-coding}. %
The resulting tags are shown in the Appendix \cite{concepts-to-teach-conv-ai-appendix}. %
Krippendorff's Alpha between all three researchers was $\alpha \geq .800$. %
The tagged data were aggregated by union between researchers, and organized with respect to the following categories: WEIRD, non-WEIRD, child and parent.

To analyze responses to Likert scale questions, %
we utilized independent and paired t-tests, Mann-Whitney U tests and Wilcoxon signed-rank tests, according to the sample and normality of the data. Figures show statistical significance with star symbols (i.e., *: $p \leq .05$, **: $p \leq .01$ and ***: $p \leq .001$). The surveys are shown in %
\cite{vanbrummelen-phd}.

\section{Results}\label{sec:brave-results}
This section describes the results most relevant to pedagogical DRs. We describe other results (e.g., most relevant to agent DRs) in \cite{vanbrummelen-phd,want-perceive-trust-transparency-chi}.

\subsection{Overall Participants}
\label{sec:res-overall}

Overall participants' feelings towards CAs shifted towards more of a friend (than a co-worker) after going through the workshops (pre/post: \={x}=3.58,3.24; t(32)=2.15; p=.039).
When asked in a long-answer question why they trusted or distrusted CAs to provide correct information, they generally provided answers for why they distrusted CAs, both before (72\%) and after (64\%) the programming activity.

\begin{figure}[htb!]
       \centering
	\includegraphics[width=\linewidth]{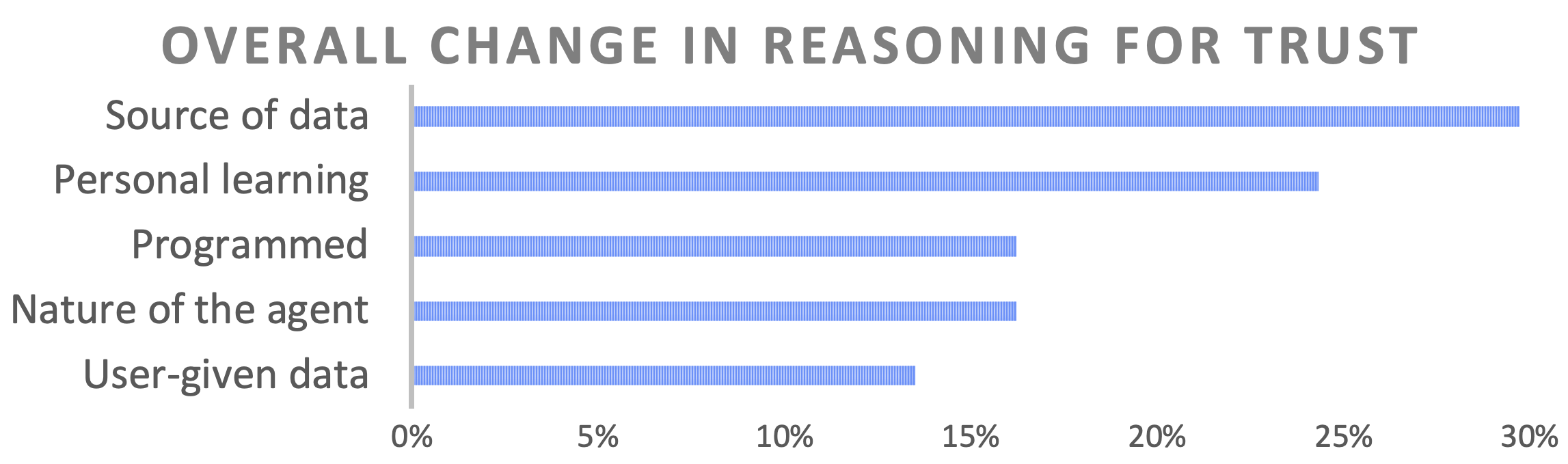}
	\caption{Participants' reasoning for changes in trust of CAs in terms of percent tag frequency. The Appendix \protect{\cite{concepts-to-teach-conv-ai-appendix}} provides descriptions of each tag.}
	\label{fig:trust-changed-reasoning-overall}
\end{figure}

About a quarter (24\%) of participants' long-answer responses indicated they felt their trust of CAs changed through the programming activity. %
As shown in Figure \ref{fig:trust-changed-reasoning-overall}, participants most often cited the source of the CAs' information (including human data, the internet and other sources) as the reason for their opinion changing. These answers, as well as those from the ``Programmed'' and ``Personal learning'' categories generally alluded to how participants had learned something through the workshops. %

In terms of \citeauthor{trust-e-commerce-typology-model}'s trust model, overall participants most often cited predictability and did not cite benevolence when discussing changes in trust (see Table \ref{tab:long-ans-trust-changed-model}). %

\begin{table}[htb!]
\centering
\caption{Percent of long-answer responses indicating different aspects of %
trust when participants discussed changes in their trust of CAs through the programming activity.}
\label{tab:long-ans-trust-changed-model}
\begin{tabular}{l|llll}
Subset & C\footnotemark & I & P & B \\ 
\hline
Overall & 32\% & 29\% & \textbf{39\%} & 0\% \\
Non-WEIRD & \textbf{40\%} & 20\% & \textbf{40\%} & 0\% \\
WEIRD & 28\% & 33\% & \textbf{39\%} & 0\% \\
Child & 13\% & \textbf{47\%} & 40\% & 0\% \\
Parent & \textbf{54\%} & 8\% & 38\% & 0\%
\end{tabular}
\\$^1$\raggedright\footnotesize C: Competence, I: Integrity, P: Predictability, B: Benevolence
\end{table}

Participants indicated \textit{Training}, CA terminology and \textit{Turn-taking} as the three most difficult things to learn in the workshops. In Figure \ref{fig:difficult-concepts-weird-non-weird} and \ref{fig:difficult-concepts-parents-children} the concepts are ordered from most to least challenging, as chosen by overall participants.
\subsection{WEIRD vs. Non-WEIRD}
\begin{figure}[htb!]
       \centering
	\includegraphics[width=.9\linewidth]{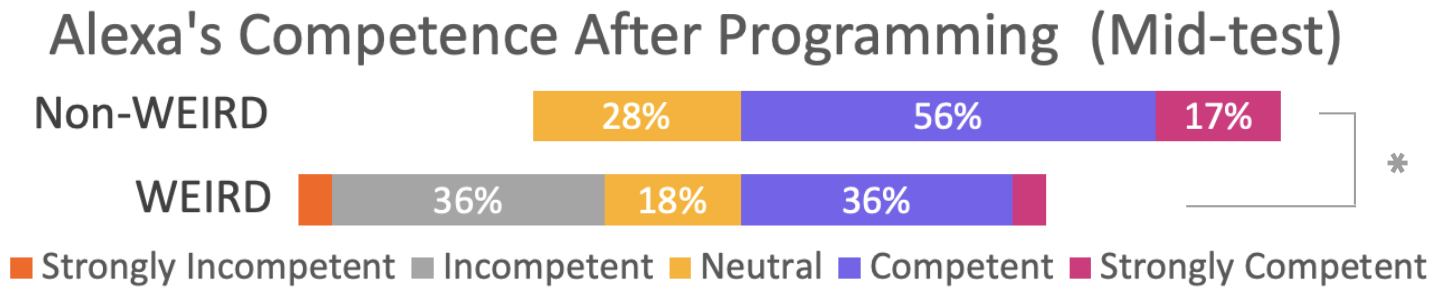}
	\caption{Distribution of responses on a 5-point Likert scale question given after the programming activity from participants from non-WEIRD and WEIRD countries when asked to rate Alexa's competence.}
	\label{fig:weird-non-weird-competence-mid}
\end{figure}

Participants from WEIRD countries thought Alexa was less competent after the programming activity than before (pre/mid: \={x}=2.43,2.95; t(20)=-2.33; p=.030). This resulted in a significant difference between participants from WEIRD and non-WEIRD countries' feelings about Alexa's competence after the programming activity (\={x}=3.00,2.11; U(38)=106.5; p=.004), as shown in Figure \ref{fig:weird-non-weird-competence-mid}. 

\begin{figure}[htb!]
       \centering
	\includegraphics[width=\linewidth]{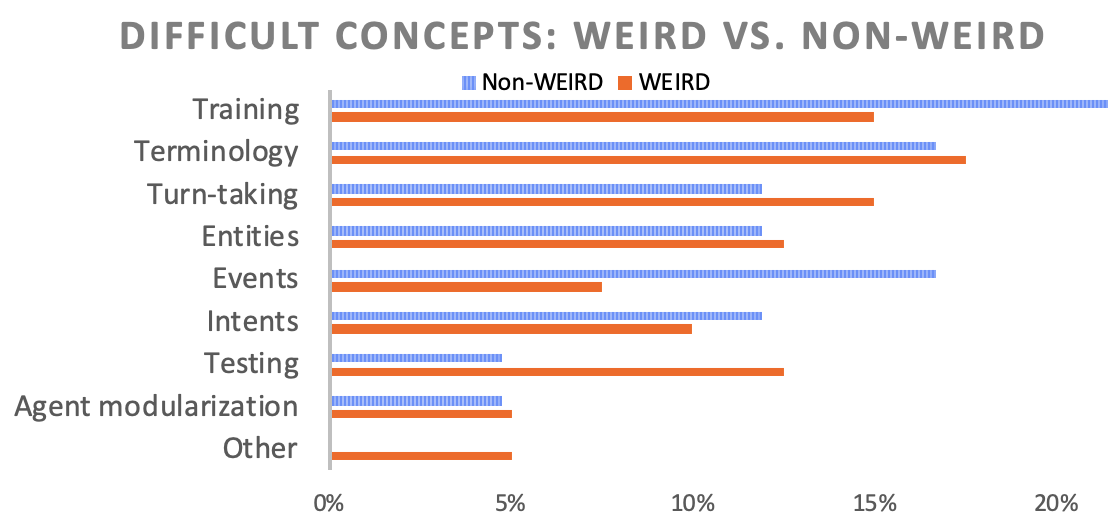}
	\caption{Relative frequency of participants from WEIRD vs. non-WEIRD countries' responses to a question asking which concepts were most difficult to learn.} %
	\label{fig:difficult-concepts-weird-non-weird}
\end{figure}

Those from non-WEIRD countries more often cited \textit{Training} and \textit{Events} as difficult concepts than those from WEIRD countries; whereas those from WEIRD countries more often cited \textit{Testing} and \textit{Turn-taking}, and more often described other concepts. Otherwise, the relative frequencies were quite similar, as shown in Figure \ref{fig:difficult-concepts-weird-non-weird}.

\subsection{Parents vs. Children}

Children had more prior experience programming than parents (\={x}=1.59,0.79; U(44)=144.5; p=.0026). Children from non-WEIRD countries had more prior experience learning about AI (\={x}=0.62,0.13; U(19)=27; p=.017) than parents from non-WEIRD countries (although this was not so for those from WEIRD countries). Children from WEIRD countries completed %
more tutorials than parents from WEIRD countries (\={x}=2.14,1.14; U(16)=18.5; p=.0.025).

\begin{figure}[htb!]
       \centering
	\includegraphics[width=\linewidth]{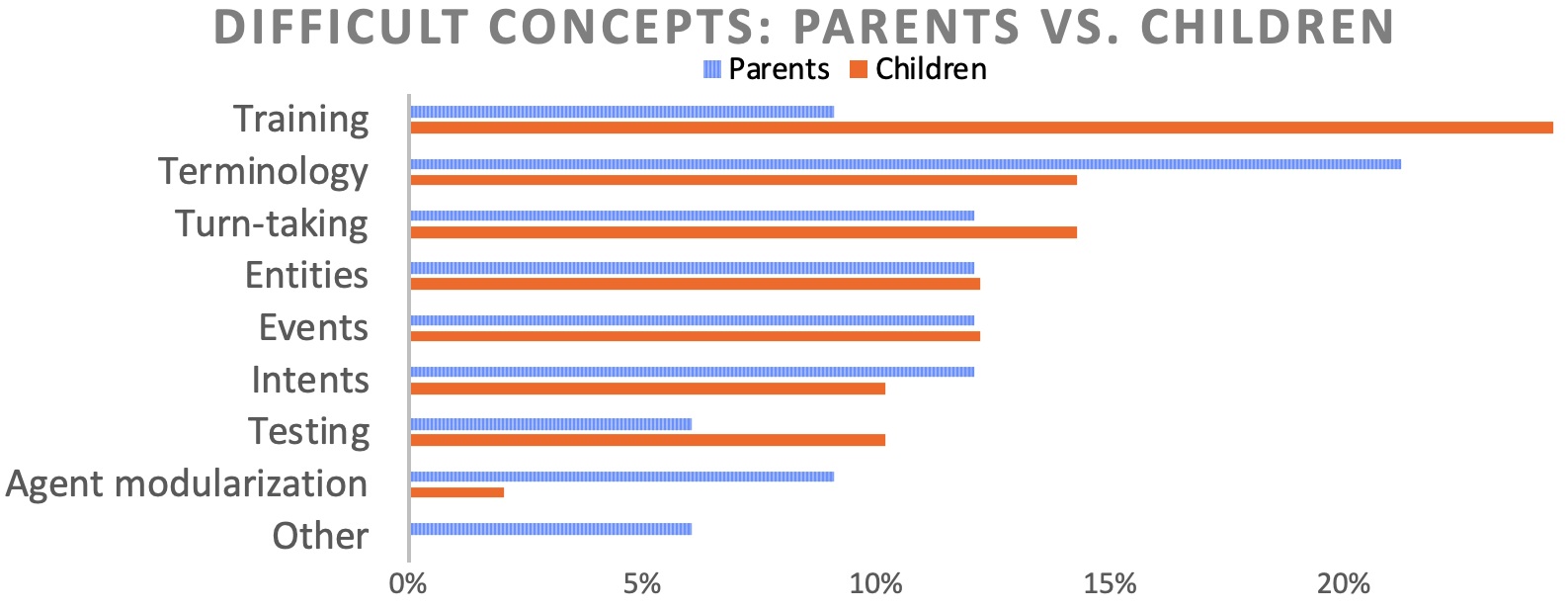}
	\caption{Relative frequency of children and parents' responses to %
	which concepts were most difficult to learn.} %
	\label{fig:difficult-concepts-parents-children}
\end{figure}

Children more often cited \textit{Training} and \textit{Testing} as difficult concepts than parents did; whereas parents more often cited CA terminology and \textit{Agent modularization}, and more often described other concepts. Otherwise, the relative frequencies were quite similar, as shown in Figure \ref{fig:difficult-concepts-parents-children}.

\subsection{Different Levels of Prior Programming Experience}
According to the pre-survey, 30\% of the participants had no prior programming experience, 13\% had only visual/block-based programming experience, and 57\% had text-based programming experience. %
There were no significant differences found in the number of tutorials completed based on participants' prior programming experience.

Prior to the workshops, those who had text-based programming experience thought Alexa was less competent (\={x}=2.73,2.07; W(16)=0; p=.038) than those who had no programming experience.
Prior to (\={x}=3.54,1.86; U(38)=62; p=\num{2.50e-4}), during (\={x}=3.86,2.25; U(31)=46; p=.0011), and after (\={x}=3.84,2.50; U(25)=30; p=.0065) the workshops, those who had text-based programming experience saw themselves more as programmers than those who had no experience initially. Prior to the workshops, those who had text-based programming experience also saw themselves more as programmers than those who had only visual programming experience (\={x}=3.54,2.33; U(30)=37; p=.022). %
After the workshops, however, there was no significant difference between how participants with text-based vs. visual programming experience felt as programmers.

\subsection{Different Prior Experience Learning about AI}
On the pre-survey, 46\% of the participants reported having no prior experience learning about AI, whereas 54\% reported they had. There were no significant differences found in the number of tutorials completed depending on whether participants had previously learned about AI or not.

Prior to the workshops, those who had learned about AI previously thought Alexa was more human-like (\={x}=2.05,2.84; U(44)=-2.87; p=.0063). Participants who had not learned AI before thought Alexa was more dependable after the programming experience (pre/mid: \={x}=3.47,3.88; W(16)=0; p=.020).
Those who had learned about AI previously saw themselves more as programmers than those who had not prior to (\={x}=2.29,3.36; U(44)=-2.73; p=.0092), during (\={x}=2.82,3.50; U(37)=129; p=.047) and after (\={x}=3.83,2.87; U(31)=69; p=.0073) the workshops. They were also more confident they could design and create their own technology project than those who did not have prior AI experience prior to (\={x}=2.71,3.60; U(44)=-2.51; p=.016), during (\={x}=3.12,4.00; U(37)=92.5; p=.0026), and after (\={x}=4.28,3.13; U(31)=44.5; p=\num{2.7e-4}) the workshops. They were also more confident they could make an impact in their community or the world using technology than those who did not have prior AI experience prior to (\={x}=3.29,4.00; U(44)=-2.52; p=.015), during (\={x}=3.18,4.18; U(37)=86.5; p=.0015), and after (\={x}=4.22,3.60; U(31)=80; p=.018) the workshops.

\begin{figure}[htb!]
       \centering
	\includegraphics[width=.65\linewidth]{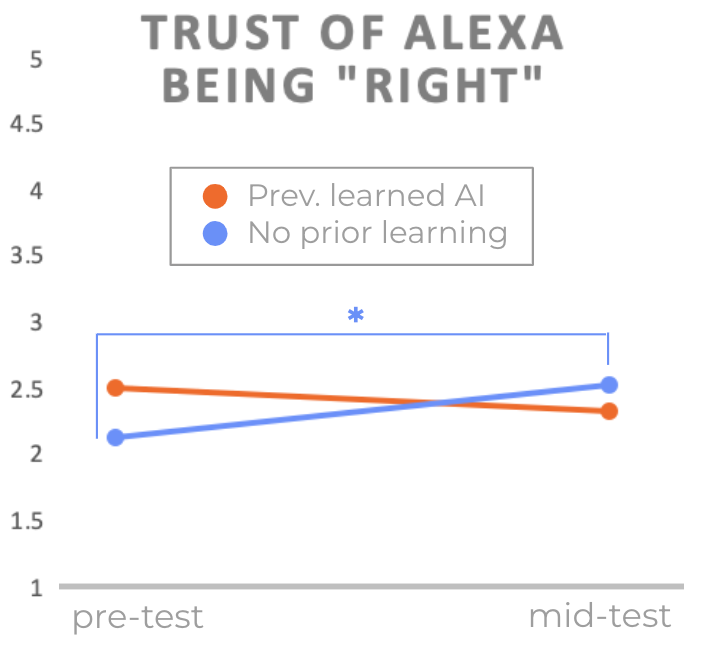}
	\caption{Mean responses to a 5-point Likert scale question about trust of Alexa's correctness given before/after the programming activity from participants with no experience and prior experience learning about AI.}
	\label{fig:no-and-yes-prior-ai-trust}
\end{figure}

\begin{figure}[htb!]
       \centering
	\includegraphics[width=\linewidth]{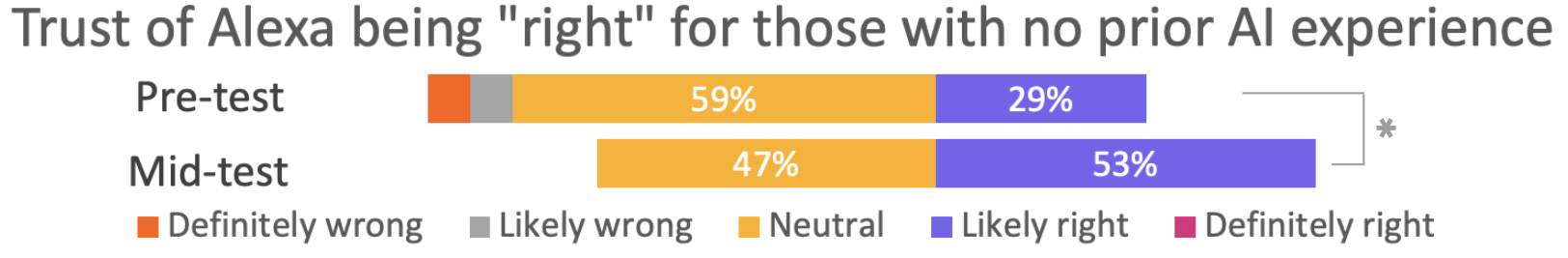}
	\caption{Distribution of responses from a 5-point Likert scale question about trust of Alexa's correctness given before and after the programming activity from participants with no prior experience learning about AI.}
	\label{fig:no-prior-ai-trust}
\end{figure}

Those who had never learned about AI before felt CAs reported correct information more after the programming activity than before (pre/mid: \={x}=2.88,2.47; W(16)=0; p=.038). No significant difference was found for those who had learned about AI before. (See Figure \ref{fig:no-and-yes-prior-ai-trust} and \ref{fig:no-prior-ai-trust}.)

\subsection{Different Experiences with Conversational Agents}
According to the pre-survey, 52\% of the participants had used more than one type of CA, 35\% had used only a single type of CA, and 13\% had never used a CA before. %
83\% %
reported typically using CAs in their first language and 17\% reported typically using them in another language.
There were no significant differences found in the number of tutorials completed depending on whether participants had used more than one type of CA or only a single CA, or on whether participants typically used CAs in their first language or not.

\begin{figure}[htb!]
       \centering
	\includegraphics[width=\linewidth]{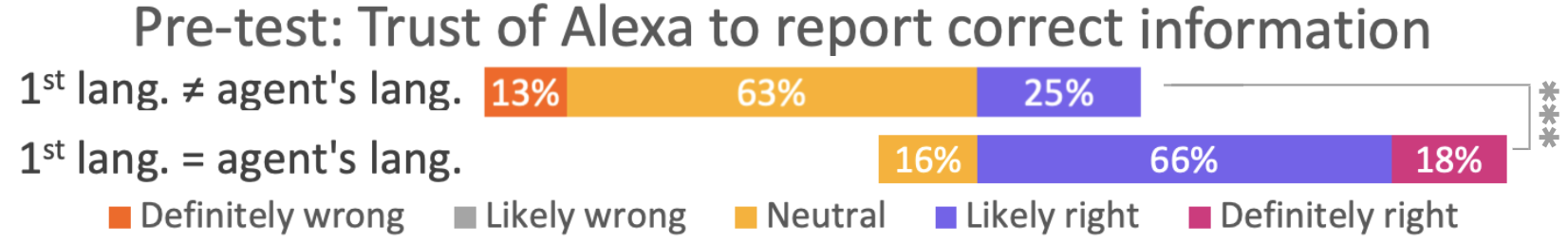}
	\caption{Distribution of responses from a 5-point Likert scale question about trust of Alexa's correctness given before the programming activity from participants who typically used CAs in their first language or not.}
	\label{fig:lang-trust-pre}
\end{figure}

\begin{figure}[htb!]
       \centering
	\includegraphics[width=.95\linewidth]{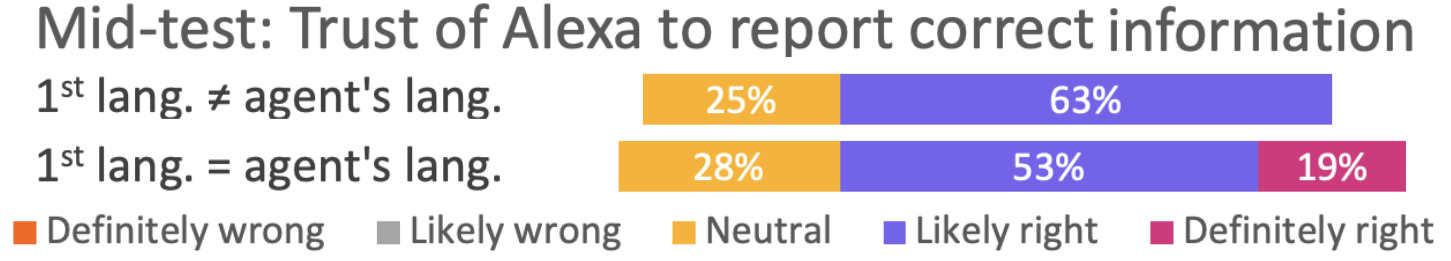}
	\caption{Distribution of responses from a 5-point Likert scale question about trust of Alexa's correctness given after the programming activity from participants who typically used CAs in their first language or not.}
	\label{fig:lang-trust-mid}
\end{figure}

Those who used CAs in their first language thought Alexa was more human-like prior to (\={x}=2.61,1.88; U(44)=87.5; p=.025), during (\={x}=2.66,2.00; U(37)=67; p=.042) and after (\={x}=2.82,1.80; U(31)=32; p=.025) the workshops than those who used them in another language. They also thought Alexa was more correct than those who used it in another language, prior to the workshops (\={x}=4.03,3.00; U(44)=52; p=\num{5.50E-4}).
As shown in Figure \ref{fig:lang-trust-pre}, prior to the workshops, participants who typically used CAs in their first language thought Alexa was more correct (\={x}=4.03,3.00; U(44)=52; p=\num{5.51e-4}) than those who typically used it in another language. There was no significant difference after the programming activity, as shown in Figure \ref{fig:lang-trust-mid}.

\section{Discussion}\label{sec:brave-discuss}
In this section, we develop DRs for teaching conversational AI by discussing the results with respect to our RQs.

\subsection{RQ1: Difficulties Learning about CAs}\label{sec:brave-discuss-learning}

With respect to RQ1, the top three concepts most referenced as difficult in this study were \textit{Training}, CA terminology and \textit{Turn-taking}. Although this trend remained relatively similar for all major subsets, different subsets of the participants still found different concepts more challenging than others. %
For example, children cited \textit{Training} and \textit{Testing} more often as difficult concepts than parents did; whereas parents cited CA terminology and \textit{Agent modularization} more often than children did. Participants from non-WEIRD countries cited \textit{Training} and \textit{Events} more often as difficult concepts than those from WEIRD countries did; whereas participants from WEIRD countries cited \textit{Testing} and \textit{Turn-taking} more often than those from non-WEIRD countries did. %
In other studies, students found \textit{Constrained vs. unconstrained natural language}, \textit{Machine learning} and \textit{Societal impact and ethics} particularly challenging to learn \cite{convo-ml,teaching-tech-talk-eaai}. Educators may want to focus on particularly challenging concepts for their students; thus, we propose, \textbf{DR1: Emphasize concepts that are challenging for particular audiences}.

\noindent\fbox{%
    \parbox{\linewidth}{%
    Opportunities to implement \textbf{DR1:} %
    \begin{itemize}
        \item Emphasize \textit{Training}, \textit{Turn-taking}, \textit{Machine learning}, \textit{Societal impact and ethics}, \textit{Constrained vs. unconstrained natural language}, and CA terminology when teaching CA curricula
        \item When teaching children, emphasize \textit{Training} and \textit{Testing}
        \item When teaching parents, emphasize CA terminology and \textit{Agent modularization}
        \item When teaching those from non-WEIRD countries, emphasize \textit{Training} and \textit{Events}
        \item When teaching those from WEIRD countries, emphasize \textit{Testing} and \textit{Turn-taking}
    \end{itemize}
    }%
}

\subsection{RQ2: Self-Efficacy and Identity as Programmers}\label{sec:brave-discuss-self-efficacy-identity}
In our study, those with additional experience in related topics generally had increased self-efficacy and identified more as programmers. For instance, those with previous AI experience identified more as being able to design and create their own technology projects, being programmers and being able to make impacts in their communities %
using technology throughout the intervention. Those with experience with more than one type of CA (as opposed to a single CA) felt similar increases as those who had learned AI. Those with text-based programming experience saw themselves more as programmers than those with no prior experience throughout the intervention. For those with initial visual programming experience, however, after engaging with the CA workshops, there was no significant difference in terms of programming identity with those who had text-based experience. Thus, creating meaningful visual programming projects and engaging in societal impact curricula may impact those with some visual programming experience more than those without any. Thus, we propose, \textbf{DR2: Supplement CA development activities with additional CA engagement, AI learning and programming activities}, 
as these opportunities likely significantly affect people's identity and self-efficacy as programmers. Also note, however, how these prior experiences (with additional CAs, AI learning activities and programming) may indicate differences in socioeconomic class, so the experiences themselves may not have caused the benefits seen in this study. Nonetheless, providing more opportunities for these types of activities to diverse audiences is key to democratizing technology.

\noindent\fbox{%
    \parbox{\linewidth}{%
        Opportunities to implement \textbf{DR2:}%
        \begin{itemize}
            \item Encourage activities such as AI learning experiences, experiences with more types of CAs and programming, in addition to CA curricula activities, as the additional activities may increase self-efficacy and identification as programmers, and lead to better CA learning outcomes
        \end{itemize}
    }
}

\subsection{RQ3: Partner Models and Trust}\label{sec:brave-discuss-partner-models-trust}

Through interacting more with Alexa and going through the workshops, participants overall felt Alexa was more of a friend. %
Such increased feelings of friendship may also increase feelings of trust long-term \cite{friendship-trust-philosophy,trust-anthropomorphism-relationships-cas}. Furthermore, those without prior AI knowledge trusted agents more after learning to program them. Considering %
how trust and student-teacher relationships are important factors when learning \cite{avatar-digital-learning-schobel,socioemotional-closeness-attachment-al-yagon}, %
it may be helpful for CAs in teaching roles to be personified. %
This trust-building may be encouraged by teachers, through increased interactions with CAs, or through activities in which students program CAs, depending on the student audience.

Nonetheless, this trust-building could lead to over-trust of agents, which can have serious consequences \cite{misinformation-trust-conspiracy-beliefs,trust-anthropomorphism-relationships-cas}. In our study, participants trusted Alexa more than their friends or parents in terms of information correctness (which may indicate over-trust, although we leave this question for future research). Thus, we propose, \textbf{DR3: Design learning activities to foster appropriate trust of agents}---which could mean either encouraging increases or decreases in trust depending on the situation.
Fortunately, students' trust towards agents was not static, and about a quarter of them indicated they felt their trust changed through the programming activities. When describing how their trust changed, participants most often referenced predictability, then competence and then integrity. They also emphasized how learning about CAs, including learning about how CAs are programmed, CAs' sources of information and how CAs understand information given to them, affected their sense of trust. Educators may want to emphasize these concepts during learning activities to encourage appropriate levels of trust.

\noindent\fbox{%
    \parbox{\linewidth}{%
        Opportunities to implement \textbf{DR3:} %
        \begin{itemize}
            \item Encourage student trust of pedagogical agents through enabling students to interact with CAs more often and in ways that encourage friendship-building
            \item Encourage student trust of pedagogical agents through teaching students how CAs work
            \item Encourage student trust of pedagogical agents through using personified or friendly pedagogical agents
            \item Encourage student reflection on agent trustworthiness by teaching about the aspects of agents' predictability, then competence and then integrity %
            \item Encourage student reflection on agent trustworthiness by teaching about how CAs are programmed, CAs' sources of information and how CAs understand information %
        \end{itemize}
    }
}

We also found different groups of participants' partner models changed differently through the activities. For instance, after the programming activity we found participants from WEIRD countries felt Alexa was less competent than those from non-WEIRD countries did. Before the workshops, we found participants without text-based programming experience thought Alexa was more competent than those with experience thought. Those with no prior AI experience thought Alexa was more dependable after learning how to program it. Throughout the workshops, participants who used CAs in their first language thought Alexa was more human-like than those who used them in another language.

Since students' partner models could affect their understanding as well as how they interact with agents \cite{partner-models-doyle}, it is important for students to have accurate partner models. Thus, we propose, \textbf{DR4: Foster accurate partner models through teaching related CA ideas and activities}. 
For example, to reinforce how CA technology is still in its infancy---and how CAs are not highly \textit{competent} in all tasks---for a WEIRD audience, a programming activity may be appropriate, but for a non-WEIRD audience, a more direct instruction approach may be appropriate. To level-set perception of CA \textit{competence} between those with and without text-based programming experience, a visual programming tutorial on CA development may be appropriate. %
To increase perceptions of CA \textit{dependability} for those who have not learned about AI before, a programming activity may be appropriate. %
To increase perceptions of CA \textit{human-likeness}, using diverse, relatable CAs %
and CAs in the audience's first language may be appropriate. %

\noindent\fbox{%
    \parbox{\linewidth}{%
        Opportunities to implement \textbf{DR4:} %
        \begin{itemize}
            \item To reinforce how CAs are not highly competent in all tasks, for a WEIRD audience, a programming activity may be appropriate, but for a non-WEIRD audience, a more direct instruction approach may be appropriate
            \item To level-set perceptions of CA competence between those with and without text-based programming experience, a visual programming tutorial on CA development may be appropriate
            \item To increase perceptions of CA dependability (if the CA \textit{is} dependable) for those who have not learned about AI before, a programming activity may be appropriate
            \item To increase perceptions of CA human-likeness (if the CA \textit{is} human-like), using diverse CAs and CAs in the audience's first language may be appropriate
        \end{itemize}
    }
}

\section{Limitations and Future Work}
While this study successfully showed various groups of people's perceptions and trust of CAs changed through learning about CAs, there were some limitations. One limitation included how there were many Indonesians in the non-WEIRD category and many Americans in the WEIRD category. Furthermore, there were more child than parent participants, meaning the overall results may be skewed towards children's results. Future studies could incorporate more participants and further balance them across subsets to verify our results. Another limitation is how we only utilized Amazon's agent in this study. This agent has a default female voice, which could affect participants' views on partner models and trust. Future studies could incorporate various types of agents. Another limitation includes how we only focused on teaching a subset of the 40 CA concepts, and how we taught the workshops virtually. Future work can investigate which of the other concepts are challenging for students, and how context (e.g., virtual vs. in-person) affects student learning. A final limitation includes how we left defining accurate/appropriate levels of trust and partner models for future research.

\section{Conclusions}
This paper presents results from investigating how children and parents from non-WEIRD and WEIRD countries' partner models and trust of CAs change through learning to program CAs and societal impact curriculum. It also presents pedagogical CA concepts (see Figure \ref{fig:agent-concepts} and the appendix \cite{concepts-to-teach-conv-ai-appendix}), design recommendations for teaching such concepts, and opportunities to implement such design recommendations with various student audiences. With these CA concepts and teaching design recommendations, educators and researchers can develop curricula to prepare students for a conversational-agent-filled world.

\bibliography{aaai23}
\bibstyle{aaai}

\end{document}